\pdfoutput=1
\documentclass[aps,twoside,twocolumn,pra,superscriptaddress]{revtex4-2}
\usepackage{epsfig,amsmath,amssymb}
\usepackage{graphicx}
\usepackage{array}
\usepackage{xcolor}

\def\vecb#1{\boldsymbol{#1}}
\def\ket#1{|#1\rangle}
\def\bket#1{\bigl|#1\bigr\rangle}
\def\bra#1{\langle#1|}
\def\bbra#1{\bigl\langle#1\bigr|}
\def\scal#1#2{\langle#1|#2\rangle}

\def\matr#1#2#3{\langle#1|#2|#3\rangle}
\def\bmatr#1#2#3{\bigl\langle#1\bigl|#2\bigr|#3\bigr\rangle}
\def\abs#1{\left\lvert#1\right\rvert}

\def\ave#1{\langle#1\rangle}

\def\={\!=\!}
\def\>{\!>\!}
\def\<{\!<\!}
\def\-{\!-\!}
\def\+{\!+\!}

\def\BLamI{\vecb{\Lambda}_{\rm I}}
\def\BLamF{\vecb{\Lambda}_{\rm F}}
\def\BLam{\vecb{\Lambda}}

\def\uvo#1{\lq\lq #1\rq\rq}

\begin{document}

\title{Decoherence-assisted quantum driving}

	\author{Pavel Cejnar}
	\email{cejnar@ipnp.mff.cuni.cz}
	\affiliation{Institute of Particle and Nuclear Physics, Faculty of Mathematics and Physics, Charles University,
		V Hole{\v s}ovi{\v c}k{\' a}ch 2, 180 00 Prague, Czechia}
	\author{Pavel Str{\' a}nsk{\' y}} 
	\email{stransky@ipnp.mff.cuni.cz}
	\affiliation{Institute of Particle and Nuclear Physics, Faculty of Mathematics and Physics, Charles University,
		V Hole{\v s}ovi{\v c}k{\' a}ch 2, 180 00 Prague, Czechia}
	\author{Jan St{\v r}ele{\v c}ek}
	\email{strelecek@ipnp.mff.cuni.cz}
	\affiliation{Institute of Particle and Nuclear Physics, Faculty of Mathematics and Physics, Charles University,
		V Hole{\v s}ovi{\v c}k{\' a}ch 2, 180 00 Prague, Czechia}
	\author{Felipe Matus}
	\email{matus@ipnp.mff.cuni.cz}
	\affiliation{Institute of Particle and Nuclear Physics, Faculty of Mathematics and Physics, Charles University,
		V Hole{\v s}ovi{\v c}k{\' a}ch 2, 180 00 Prague, Czechia}

\date{\today}

\begin{abstract}
We propose a protocol for transitionless driving of a bound quantum system in its parameter space using repeated measurement-like interactions with an external spectator system.
As a consequence of the quantum Zeno effect, the fidelity of the final state preparation is equal to unity in the limit of infinite-rate interactions.
For finite-rate interactions, the maximal fidelity is achieved for the driving trajectory having a minimal geometric length and keeping a constant speed with respect to the Provost-Vallee metric in the parameter space.
We numerically test the protocol in an interacting multiqubit system, demonstrating its dominance over the method of coherent driving.  
\end{abstract}

\maketitle

{\it Introduction.---}
A possible way to build a quantum computer relies on the idea of adiabatic quantum computation \cite{Fahr00,Alba18}. 
It is formulated for an isolated many-body system with a discrete energy spectrum depending on a~set of controllable parameters $\BLam$ (external filds or internal coupling strengths).
The aim is to prepare the ground state at ${\BLam=\BLamF}$, where the system exhibits complex correlations between individual constituents (these will be exploited in a~particular quantum algorithm).
The adiabatic theorem of quantum mechanics~\cite{Born28} ensures that for a very slow variation of parameters the system remains in the instantaneous energy eigenstate.
So the desired final state can be obtained by initiating the system in the ground state at ${\BLam=\BLamI}$, where the correlations are not present, and then slowly changing the parameters along a~certain path in the parameter space to the final point~$\BLamF$.
The overall duration $T$ of the driving procedure must be sufficiently long to avoid unwanted excitations of the system during the drive.

Since the adiabaticity-violating effects may be strong even for rather slow driving \cite{Garr62,Berr87,Nenc93,Teuf03,Orti08}, the requirement of high fidelity of the driving procedure may set too stringent constrains on its overall duration, preventing suitable scaling with the size of the many-body system.
This is particularly true if the initial and final states are separated by a finite-size precursor of a quantum phase transition (QPT) \cite{Sach99,Zure05,Dams05,Schu06}.
A method that can potentially overcome this problem, the so-called counterdiabatic driving, is based on adding some extra terms into the driven Hamiltonian to compensate nonadiabatic effects \cite{Demi03,Berr09,Camp13,Sels17}.
If the energy costs connected with action of the additional terms are ignored, this strategy can in principle work for any preselected driving time~$T$. 
However, realistic energy restrictions imply lower bounds on time.
Research of general time constraints in preparation of selected quantum states and their interplay with energy relations characterizing the system constitutes a~quickly expanding field of quantum speed limits \cite{Deff17,Fran16,Buko19}.

In this paper, we propose another strategy to keep the fidelity of the driving procedure close to unity.
Our method combines sudden jumps along a selected trajectory in the parameter space with periodically performed measurement-like interactions with an external spectator system, which repeatedly diagonalize the density matrix of the controlled system in the running Hamiltonian eigenbasis.
This approach is closely related to the measurement-based techniques of quantum control \cite{Roa06,Pech06,Schu08,Haco18}.
It is known that, in analogy to the quantum Zeno effect \cite{Misr77,Itan90}, the final-state fidelity tends to unity if the interaction rate grows to infinity, and even imperfect realizations of the procedure with finite-rate interactions can provide a~significant improvement of fidelity.

A crucial question related to these finite-rate realizations is which path in the parameter space and which set of interaction points yield the best performance. 
The purpose of this Letter is to demonstrate---on the general ground as well as using a particular model example---that the answer to this question follows from the geometric description of parameter-dependent quantum systems in terms of the formalism of curved spaces \cite{Prov80,Berr84,Wilc88,Kolo17,Buko19}. 

{\it Decoherence-assisted driving protocol.---}
We assume a~general Hamiltonian $\hat{H}(\BLam)$ depending on a $D$-dimen\-sional set of parameters ${\BLam\equiv\Lambda^\mu}$, ${\mu=1,2,\dots,D}$, with normalized eigenvectors $\ket{E_i(\BLam)}$ assigned to discrete nondegenerate energy levels $E_i(\BLam)$. 
The index ${i=0,1,2,\dots}$ increases with energy, so ${i=0}$ corresponds to the ground state, ${i=1}$ to the first excited state, etc. 
The system is driven along a path $\wp\equiv\{\BLam(t)\}_{t=0}^{T}$ in the parameter space satisfying ${\BLam(0)=\BLamI}$ and ${\BLam(T)=\BLamF}$.
The dependence $\BLam(t)$ determines the shape of the trajectory as well as the course of the motion along it.

We first introduce the case of coherent driving, when the system is isolated and remains in a pure state $\ket{\psi(t)}$ determined by the Schr{\"o}dinger equation with a time-dependent Hamiltonian $\hat{H}(\BLam(t))$. 
Using an expansion ${\ket{\psi(t)}=\sum_i\alpha_i(t)\ket{E_i(\BLam(t))}}$, where $\alpha_i(t)$ are normalized amplitudes, we express matrix elements of the density operator ${\hat{\varrho}^{\rm coh}(t)=\ket{\psi(t)}\bra{\psi(t)}}$ of the coherent evolution in the instantaneous Hamiltonian eigenbasis:
\begin{equation}
\varrho^{\rm coh}_{ij}(t)\=\bmatr{E_i(\BLam(t))}{{\hat\varrho}^{\rm coh}(t)}{E_j(\BLam(t))}\=\alpha_i(t)\alpha_j^*(t).
\label{nodia}
\end{equation}
We suppose that the system is initiated in the ground state $\ket{E_0(\BLamI)}$ and that the purpose of our driving procedure is to prepare the ground state $\ket{E_0(\BLamF)}$.
The fidelity of the instantaneous ground state at any intermediate time ${t\in[0,T]}$ reads ${{\cal F}^{\rm coh}(t)=\varrho^{\rm coh}_{00}(t)}$.
The adjunct ${{\cal I}=1-{\cal F}}$ of fidelity ${\cal F}$ will be called infidelity.

Density matrix \eqref{nodia} is generically nondiagonal due to coherence of the isolated system evolution.
Below we assume that the coherence is destroyed by measurement-like interactions that repeatedly erase off-diagonal elements of the density matrix in the running Hamiltonian eigenbasis.
So we consider an alternative evolution 
\begin{equation}
\varrho_{ij}(t)=p_i(t)\delta_{ij},
\label{dia}
\end{equation}
where $p_i(t)$ are normalized occupation probabilities that generally differ from $\varrho_{ii}^{\rm coh}(t)$ of the coherent evolution. 
The instantaneous fidelity (infidelity) is given by
\begin{equation}
{\cal F}(t)=\varrho_{00}(t)=p_0(t)=1-{\cal I}(t).
\label{fidel2}
\end{equation}

\begin{figure}[tp]
	\includegraphics[width=0.85\linewidth]{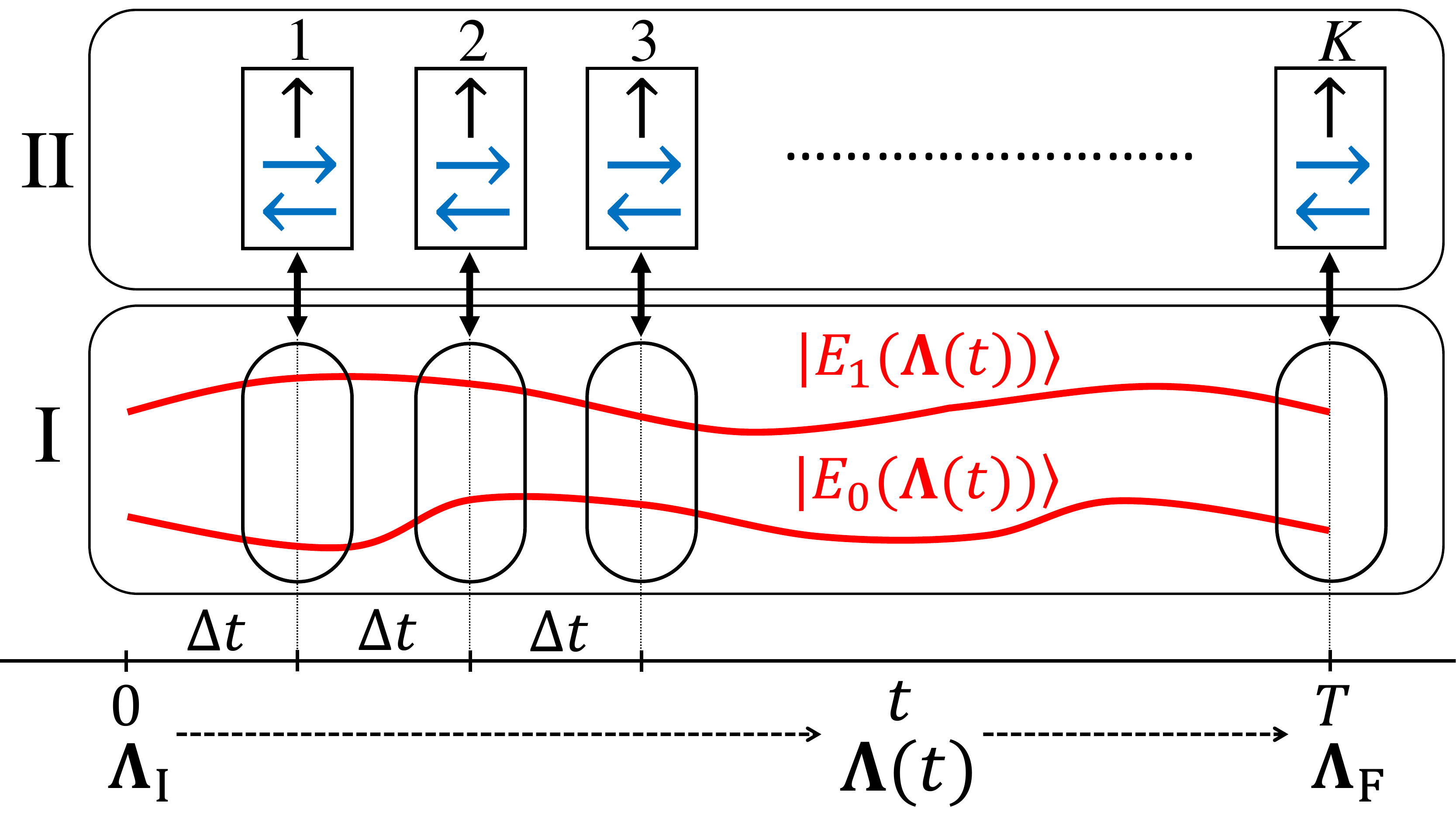}
	\caption{
A sketch of the decoherence-assisted driving for a~two-level system.
The driven system~I with a parameter-dependent eigenvectors $\bket{E_i(\BLam)}$, ${i=0,1}$ is repeatedly coupled to a spectator spin-$\frac{1}{2}$ system~{II} initialized at times ${t=t_k}$ with ${k=1,2,\dots,K}$ in the \uvo{up} state $\ket{\!\uparrow}$.
An interaction in the interval ${t\in(t_k,t_k+\Delta t)}$ generates an entangled state of the whole system such that the eigenstate $\bket{E_0\bigl(\BLam(t_k)\bigr)}$ is coupled to $\ket{\!\leftarrow}$ and $\bket{E_1\bigl(\BLam(t_k)\bigr)}$ to $\ket{\!\rightarrow}$, where the arrows denote spin states with the respective orientations satisfying ${\scal{\leftarrow\!\!}{\!\!\rightarrow}=0}$.
The resulting full decoherence of system~I is equivalent to an ideal quantum measurement in the basis $\ket{E_i(\BLam)}$.
The procedure can be realized with $K$ different spins employed consecutively at each coupling event, or with a single spin which is externally reset at each $t_k$ to the initial state $\ket{\!\uparrow}$. 
	}
	\label{sketch}
\end{figure}

The decoherence process can be realized either via a~repeated quantum measurement on the driven system, or via its specific interaction with an external spectator system.
The first option assumes a collapse of the state vector in the local Hamiltonian eigenbasis, which at each measurement instant $t_k$ leads to a nonunitary transition 
\begin{equation}
\sum_i a_i\bket{E_i(\BLam_k)}
\stackrel{\Delta t}{\longrightarrow}
\sum_i|a_i|^2\bket{E_i(\BLam_k)}\bbra{E_i(\BLam_k)}
\label{trans1}
\end{equation}
from a pure state (left) to a mixed state (right).
Here,~$a_i$ denote normalized amplitudes resulting from the evolution after the previous measurement, ${\BLam_k\equiv\BLam(t_k)}$, and~$\Delta t$ represents a time scale of the measurement procedure. 
The second realization of the decoherence process counts on interaction with an external spectator system which induces a~unitary evolution 
\begin{equation}
\bigl[\,\sum_i a_i\bket{E_i(\BLam_k)}\bigr]\!\otimes\!\ket{S}
\stackrel{\Delta t}{\longrightarrow}
\sum_i a_i\bket{E_i(\BLam_k)}\!\otimes\!\ket{S_i}
\label{trans2}
\end{equation}
from a factorized state, where $\ket{S}$ stands for an initial state of the spectator, to an entangled state involving spectator states $\ket{S_i}$ that satisfy the orthogonality relation ${\scal{S_i}{S_j}=\delta_{ij}}$.
The partial density operator of the driven system extracted from the right-hand side of Eq.\,\eqref{trans2} coincides with that in Eq.\,\eqref{trans1}, so both methods are equivalent. 
A sketch of the decohering procedure for a two-level system is shown in Fig.\,\ref{sketch}.

In the following, the decoherence-assisted driving protocol is implemented as a discontinuous procedure which consists of a sequence of small quenches---sudden jumps between discrete parameter points that form a finite sample of the selected path~$\wp$.
The total driving time~$T$ is split into $K$ equal intervals ${\Delta t=T/K}$ and at each time ${t_k=k\Delta t}$, ${k=0,1,\dots,K}$ the parameters are abruptly switched from $\BLam_k$ to $\BLam_{k+1}$.
This stroboscopic procedure can nevertheless be treated as a quasicontinuous driving since the number~$K$ is assumed to be very large, so the discretization of the path~$\wp$ is rather fine-grained.

The waiting time $\Delta t$ at each parameter point $\BLam_k$ with ${k>0}$ is set to a time $\tau$ needed for the transition \eqref{trans2} to the fully decohered density matrix.
This time is determined by the system-spectator interaction (an example will be discussed below) and we assume that it is independent of $\BLam$.
Since the condition ${\Delta t=\tau}$ should be satisfied sharply, the total driving time is ${T=K\tau}$.

{\it The optimal path.---}
Due to the measurement-like interactions, the above stroboscopic driving procedure can be described as a classical Markovian probabilistic process.  
The population probability for the $i$th Hamiltonian eigenstate after the $k$th quench reads
\begin{eqnarray}
p_i(t_{k})&=&\sum_{i'}B_{ii'}(t_{k-1})p_{i'}(t_{k-1}),
\label{recur}\\
B_{ii'}(t_{k-1})&=&\abs{\scal{E_i(\BLam_{k})}{E_{i'}(\BLam_{k-1})}}^2,
\nonumber
\end{eqnarray}
where $p_{i'}(t_{k-1})$ are population probabilities in the previous step and $B_{ii'}(t_{k-1})$ are branching ratios for ${i'\to i}$ transitions in the $k$th quench. 
At any fixed time we have ${\sum_{i}B_{ii'}=1}$ and ${\sum_{i'}B_{ii'}=1}$.
Equation~\eqref{recur} enables us to determine all probabilities $p_i(t_k)$ with ${k>0}$ recursively from the values 
$p_i(t_0)$ at the initial time.
In our case, the initial condition reads ${p_i(t_0)=\delta_{i0}}$.

We will calculate the ground-state fidelity ${\cal F}(t)$ in the running discrete time.
If the ground state is populated at ${t=t_k}$ and if the length ${[\sum_\mu(\delta_k\Lambda^{\mu})^2]^{1/2}}$ of the subsequent quench ${\BLam_k\to\BLam_{k+1}}$ with ${\delta_k\Lambda^{\mu}\equiv\Lambda^{\mu}_{k+1}\!-\Lambda^{\mu}_{k}}$ is small enough, the transition to the neighboring ground state dominates over those to excited states.
The corresponding branching ratio is
\begin{eqnarray}
B_{00}(t_k)&=&1-\sum_{i>0}B_{i0}(t_k)=1-\delta\ell_k^2,
\label{gg}\\
\delta\ell_k^2&\approx&{\rm Re}\sum_{i>0}\!\!\frac{\matr{E_0}{\partial_{\mu}\hat{H}}{E_i}\matr{E_i}{\partial_{\nu}\hat{H}}{E_0}}{|E_i-E_0|^2}\delta_k\Lambda^{\mu}\delta_k\Lambda^{\nu},
\nonumber
\end{eqnarray}
where the second line with ${\partial_\mu\hat{H}(\BLam)\equiv\frac{\partial}{\partial\Lambda^{\mu}}\hat{H}(\BLam)}$ follows from the elementary perturbation theory (all quantities are implicitly taken at ${\BLam=\BLam_k}$).
Here and below we use the summation convention for indices $\mu$ and $\nu$.
For the protocols that consist of a large number of short quenches the final fidelity can be approximated by
\begin{equation}
{\cal F}(T)\approx\prod_{k=0}^{K-1}B_{00}(t_k)\approx\prod_{k=0}^{K-1}\left(1-\delta\ell_k^2\right).
\label{product}
\end{equation}
Indeed, for ${\delta\ell_k\sim{\cal O}(K^{-1})}$, see the second line of Eq.\,\eqref{gg}, the deviation of ${\cal F}(T)$ from unity is of order ${\cal O}(K^{-1})$, while a summed contribution of all transition sequences not included in Eq.\,\eqref{product} is of order ${\cal O}(K^{-2})$.

We intend to maximize the final fidelity ${\cal F}(T)$ over the set of all paths $\wp$ with fixed $\BLamI$ and $\BLamF$ and their arbitrary discretizations $\{\BLam_k\}_{k=0}^{K}$.
The solution of this problem can be found with the aid of the geometric description of parameter-dependent quantum systems in terms of the Provost-Vallee metric.
It is defined by
\begin{equation}
d\ell^2=g_{\mu\nu}(\BLam)d\Lambda^{\mu}d\Lambda^{\nu}\equiv 1-\abs{\scal{E_0(\BLam\!+\!d\BLam)}{E_0(\BLam)}}^2,
\label{metric}
\end{equation}
where $g_{\mu\nu}(\BLam)$ is a metric tensor in the space of $\BLam$ induced by the changing structure of ground state 
(the metric is independent of local gauge transformations).
We see that the element of distance $d\ell$ in Eq.\,\eqref{metric} is an infinitesimal form of quantities $\delta\ell_k$ in Eq.\,\eqref{gg}.
Therefore, all sequences $\{\delta\ell_k\}_{k=1}^{K}$ assigned to different discretizations of the same trajectory in the parameter space sum up to a roughly constant value approximating the length of the trajectory 
\begin{equation}
\ell_\wp=\int_0^T dt\,\sqrt{g_{\mu\nu}(\BLam(t))\,\dot{\Lambda}^\mu\dot{\Lambda}^\nu},
\label{length}
\end{equation}
where dots denote time derivatives.

According to Eq.\,\eqref{product}, the fidelity resulting from partitioning of the length $\ell_\wp$ by any sequence $\{\delta\ell_k\}_{k=1}^{K}$ is for large enough $K$ given by ${{\cal F}(T)\approx 1-\sum_k\delta\ell^2_k}$.
Maximal fidelity is achieved for an equidistant sequence, when the length interval passed in each step is the same, ${\delta\ell_k=\ell_{\wp}/K\approx(\sum_k\delta\ell_k)/K}$.
This means that individual quenches should be chosen so that the quasicontinuous stroboscopic motion along the selected path keeps a constant speed ${v=(g_{\mu\nu}\dot{\Lambda}^{\mu}\dot{\Lambda}^{\nu})^{1/2}}$ on the ground-state manifold equal to ${\ell_{\wp}/T}$.
This is the first condition for an optimal driving procedure.

The second condition is now very easy to derive.
Applying Eq.\,\eqref{product} to the equidistant sequence, we obtain 
\begin{equation}
{\cal I}(T)=\frac{\ell^2_{\wp}}{K}-\frac{\ell^4_{\wp}}{2K^2}-\frac{R}{K^2},
\label{infi}
\end{equation}
where ${K=T/\tau}$.
The last term with ${R>0}$ represents the unevaluated contribution from sequences connecting the initial and final ground states via intermediate excited states.
We see that for large $K$ the optimal driving follows the path with minimal length~$\ell_{\wp}$, i.e., the geodesic connecting the initial and final parameter points.

We stress that for ${K\to\infty}$ the fidelity ${\cal F}(T)$ limits to unity for any decoherence-assisted driving procedure.
This is a direct consequence of the quantum Zeno effect, in which quantum measurements performed with infinite rate freeze the system in a motionless state \cite{Misr77,Itan90}.
For a~fixed time $T$ and length~$\ell_{\wp}$ this is achieved if the decoherence time $\tau$ drops to zero.
The above-derived optimization conditions apply in realistic situations with large but finite values of~$K$.

{\it Decohering procedure for a single qubit.---}
Let us discuss a possible implementation of the decoherence process \eqref{trans2} in a driven single-qubit system, whose sketch was presented in Fig.\,\ref{sketch}. 
In the two-level case the spectator system at each time instant ${t=t_k}$ can also be just a two-level system, and we assume that it is a single spin-$\frac{1}{2}$ particle.
Denoting the spectator spin up and down states as ${\ket{\!\uparrow}}$ and ${\ket{\!\downarrow}}$, and 
the qubit energy eigenstates $\bket{E_0\bigl(\BLam(t_k)\bigr)}$ and $\bket{E_1\bigl(\BLam(t_k)\bigr)}$ as $\ket{0}$ and $\ket{1}$, we span the Hilbert space of the coupled system by basis vectors $\ket{0}\otimes\ket{\!\uparrow}$,$\ket{0}\otimes\ket{\!\downarrow}$, $\ket{1}\otimes\ket{\!\uparrow}$ and $\ket{0}\otimes\ket{\!\downarrow}$.
In both qubit and spectator parts of the product space we introduce Pauli matrices $(\hat{\sigma}_x,\hat{\sigma}_y,\hat{\sigma}_z)$ so that the ${\left(\smallmatrix 1\\0\endsmallmatrix\right),\left(\smallmatrix 0\\1\endsmallmatrix\right)}$ basis states are associated with ${\ket{1},\ket{0}}$ and ${\ket{\!\uparrow},\ket{\!\downarrow}}$.
In all formulas, the components associated with the qubit and spectator appear on the first and second position, respectively.
 
If the coupled system after the $k$th quench starts from a separated initial state ${\ket{\Psi(t_k)}=\bigl(a_0\ket{0}+a_1\ket{1}\bigr)\otimes\ket{\!\uparrow}}$, where $a_0$ and $a_1$ are arbitrary coefficients, the full decoherence from Eq.\,\eqref{trans2} is achieved if the state after time~$\tau$ evolves to the entangled state written, e.g., in the following form: ${\ket{\Psi(t_k\+\tau)}=a_0\,\ket{0}\!\otimes\!\ket{\!\leftarrow}+a_1\,\ket{1}\!\otimes\!\ket{\!\rightarrow}}$.
Here ${\ket{\!\leftarrow}\equiv(\ket{\!\uparrow}\-i\ket{\!\downarrow})/\sqrt{2}}$ and ${\ket{\!\rightarrow}\equiv(\ket{\!\uparrow}\+i\ket{\!\downarrow})/\sqrt{2}}$ denote mutually orthogonal spin states obtained by the rotation of $\ket{\!\uparrow}$ around the $x$-axis by angle $\pm\pi/2$.
The corresponding evolution operator can be written as
\begin{equation}
\hat{U}(t')=\ket{0}\bra{0}\otimes e^{-i\frac{\pi}{4\tau}\hat{\sigma}_xt'}
+\ket{1}\bra{1}\otimes e^{+i\frac{\pi}{4\tau}\hat{\sigma}_xt'},
\end{equation}
where ${t'=t-t_k}$, hence the Hamiltonian of the system--spectator interaction reads 
\begin{equation}
\hat{H}_{\rm int}=-\frac{\pi\hbar}{4\tau}\ \hat{\sigma}_z\otimes\hat{\sigma}_x.
\label{Hint}
\end{equation}
We stress that the fully decohered qubit state is only transient---it appears at time ${t=\tau}$ and then $3\tau$, $5\tau$...

As follows from Eq.\,\eqref{Hint}, any speed-up of the driving protocol, which requires a shortening of the decoherence time $\tau$, can be achieved only through an increase of the strength of the qubit-spectator interaction.
Using the first-order term from formula \eqref{infi}, we obtain the proportionality relation ${\cal I}(T)\propto\ell_{\wp}^2\hbar/S_{\rm int}$, where ${S_{\rm int}=\int_0^Tdt\,\matr{\Psi(t)}{\hat{H}_{\rm int}}{\Psi(t)}}$ is the overall action of the interaction Hamiltonian in the evolving state $\ket{\Psi(t)}$ of the coupled system. 

{\it Test in a multiqubit system.---}
We verify the efficiency of the decoherence-assisted state preparation method in a~fully connected system of $N$ qubits.
Its Hilbert space is spanned by a factorized basis ${\bigl\{\bigotimes_{i=1}^N\ket{l^{(i)}}\bigr\}}$ with ${l\in\{1,0\}}$. 
In the $i$th qubit space we introduce Pauli matrices $(\hat{\sigma}^{(i)}_{x},\hat{\sigma}^{(i)}_{y},\hat{\sigma}^{(i)}_{z})$ written in basis states ${\ket{1^{(i)}},\ket{0^{(i)}}}$ and we also define an operator ${\hat{\kappa}^{(i)}\=(\hat{\sigma}^{(i)}_z\+1)/2}$.
The Hamiltonian reads as
\begin{eqnarray}
&&\hat{H}=
-\frac{\lambda+2\chi^2}{4}+\sum_{i}
\biggl[\biggl(\frac{1}{2}\!-\!\frac{\chi^2}{2N}\biggr)\,\hat{\sigma}^{(i)}_z-\frac{\chi}{2N}\,\hat{\sigma}^{(i)}_x\biggr]
\label{Ham}\\
&&-\frac{1}{4N}\!\sum_{i\neq j}\biggl[
\lambda\,\hat{\sigma}^{(i)}_x\hat{\sigma}^{(j)}_x
\!+\!
\chi\bigl(\hat{\sigma}^{(i)}_x\hat{\kappa}^{(j)}\!+\!\hat{\kappa}^{(i)}\hat{\sigma}^{(j)}_x\bigr)
\!+\!
\chi^2\hat{\kappa}^{(i)}\hat{\kappa}^{(j)}
\biggr],
\nonumber
\end{eqnarray}
where ${\lambda\in(-\infty,+\infty)}$ and ${\chi\in[0,\infty)}$ are two control parameters (${\equiv\BLam}$) that form a halfplane.
This model was used in our previous study of coherent driving \cite{Matu22}. 

\begin{figure}[tp]
	\includegraphics[width=\linewidth]{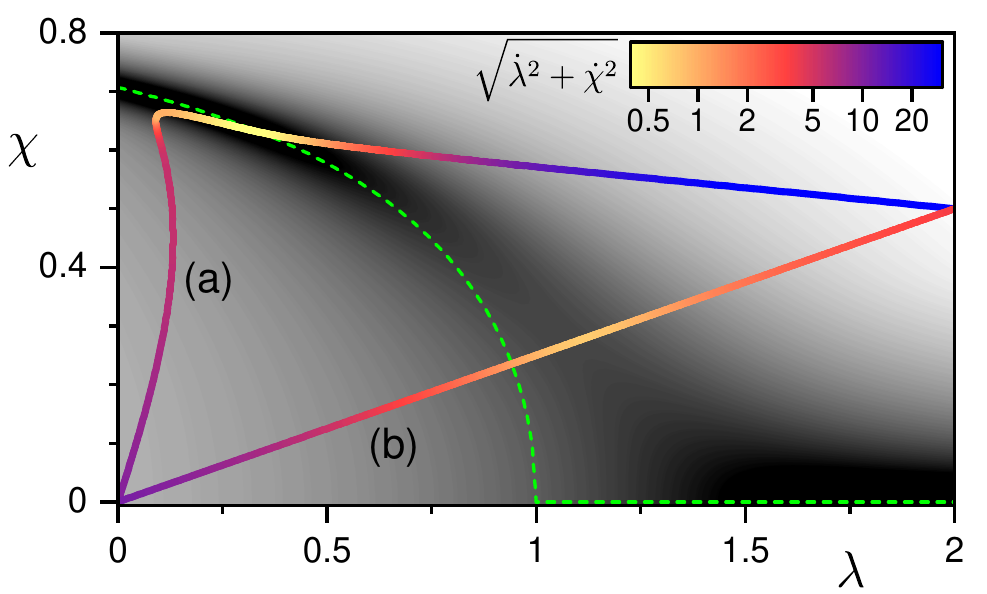}
	\caption{
(color online) Parameter plane of Lipkin Hamiltonian~\eqref{Ham} with ${N=10}$ and the geodesic (a) and linear (b) paths between the initial and final points.
The variable speed in the plane that ensures a constant speed on the ground-state manifold is expressed by the color scale (the indicated values correspond to ${T=1}$).
The gray background encodes the size of the energy gap between the ground state and first excited state (darker shades indicate smaller gaps) and the dashed curve represents the QPT separatrix in the limit ${N\to\infty}$.  
	}
	\label{geo}
\end{figure}

Hamiltonian \eqref{Ham} contains one-body terms (the sum in the first line) and two-body interactions acting equally between all pairs of qubits (the second line).
It can be rewritten in terms of collective quasispin operators ${\hat{J}_{\alpha}=\frac{1}{2}\sum_i\hat{\sigma}^{(i)}_{\alpha}}$, with ${\alpha=x,y,z}$ which casts the model as a specific version of the Lipkin model~\cite{Lipk65}.
Subspaces of the entire $2^N$-dimensional Hilbert space with different permutation symmetries related to the exchange of qubits are invariant under the evolution, so we select for our analysis the ${(N\+1)}$-dimensional fully symmetric subspace associated with the maximal value ${j=N/2}$ of the total quasispin quantum number. 

The model in its infinite-size limit has an interesting phase structure.
As the number of qubits $N$ increases, the two-body terms in Eq.\,\eqref{Ham} with ${{\cal O}(N^{-1})}$ prefactors and one-body terms with ${{\cal O}(1)}$ prefactors give comparable ${{\cal O}(N)}$ contributions to the total energy, whereas the one-body terms with ${{\cal O}(N^{-1})}$ prefactors fade away.
In the limit ${N\to\infty}$, the ground state shows two basic forms characterized by expectation values of the operators $\hat{J}_x$ and $\hat{J}_z$: a~factorized form with ${\ave{\hat{J}_x}=0}$, ${\ave{\hat{J}_z}=-j}$, so all qubits being strictly in state $\ket{0}$, and an entangled form with ${\ave{\hat{J}_x}>0}$, ${\ave{\hat{J}_z}>-j}$, so each qubit allowing for both $\ket{0}$ and $\ket{1}$ measurement outcomes.
In the halfplane ${(\lambda,\chi)}$, these ground-state phases are separated by a critical curve of the first-order QPT which ends at ${\chi=0}$ with point of the second-order QPT.
Note that a mirror symmetric structure would appear in the ${\chi<0}$ halfplane with inverted values of $\ave{\hat{J}_x}$.

\begin{figure}[tp]
	\includegraphics[width=\linewidth]{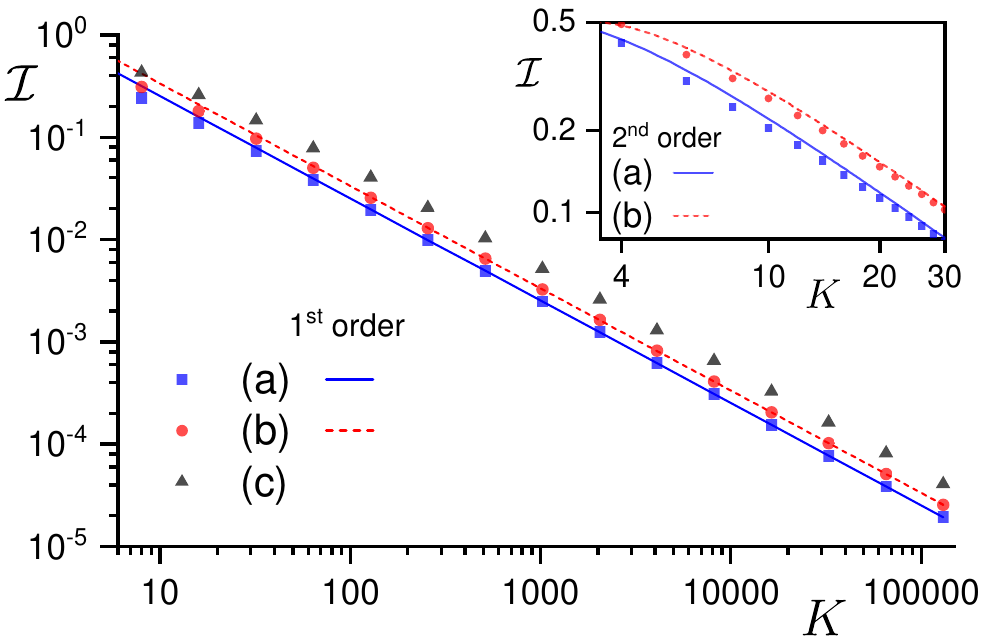}
	\caption{
Infidelity ${\cal I}(T)$ of decoherence-assisted driving protocols in the ${N=10}$ Lipkin model as a function of the number of steps $K$ for the driving paths (a), (b) and (c) described in the text (cf.\,Fig.\,\ref{geo}).
Dots represent exact values of infidelity, lines assigned to paths (a) and (b) show the approximation by formula \eqref{infi} including the first one and two terms (see the main panel and inset, respectively).
	}
	\label{lin}
\end{figure}

We test the above state-preparation method for several driving paths connecting the initial parameter point ${\BLamI\equiv(\lambda_{\rm I},\chi_{\rm I})=(0,0)}$, located in the factorized ground-state phase, with the point ${\BLamF\equiv(\lambda_{\rm F},\chi_{\rm F})=(2,0.5)}$ in the entangled phase.
We consider three types of driving: (a) the one along the geodesic trajectory with a constant speed $v$ on the ground-state manifold, (b) the driving along a linear trajectory, again with a constant speed~$v$ on the manifold, and (c) the driving along a linear trajectory with a constant speed ${u=(\dot{\lambda}^2\+\dot{\chi}^2)^{1/2}}$ in the parameter plane.
Paths (a) and (b) are shown in Fig.\,\ref{geo} with the corresponding variable speed $u$, path (c) coincides with (b) but has ${u={\rm const}}$.
All paths cross the first-order QPT separatrix (the dashed curve) as well as its finite-size realization, where the energy gap (encoded in the shades of gray) between the ground state and the first excited state exponentially decreases with $N$.

The infidelity obtained in decoherence-assisted driving protocols for the above paths~(a),\,(b) and~(c) discretized to $K$ finite steps is presented in Fig.\,\ref{lin}.
Exact values of ${\cal I}(T)$ calculated numerically are represented by dots of various shapes, while the approximations by Eq.\,\eqref{infi} for paths (a) and (b) are depicted by smooth lines [for path (c) that approximation is not valid].
We see a nice agreement of the ${\propto K^{-1}}$ term with exact results for ${K\gtrsim 50}$, while the inclusion of the first ${\propto K^{-2}}$ term makes formula \eqref{infi} applicable down to much lower values of $K$ (see the inset). 
The second ${\propto K^{-2}}$ term, which is due to transition sequences not included in Eq.\,\eqref{product}, reduces the value of ${\cal I}(T)$ for low $K$ only in a moderate way. 
The figure demonstrates dominance of the geodesic driving path (a) over the nongeodesic ones, as well as a~clear advantage of the linear path (b) with ${v={\rm const}}$ over the linear path (c) with ${u={\rm const}}$.
This all supports the theoretical analysis presented above.

\begin{figure}[tp]
	\includegraphics[width=\linewidth]{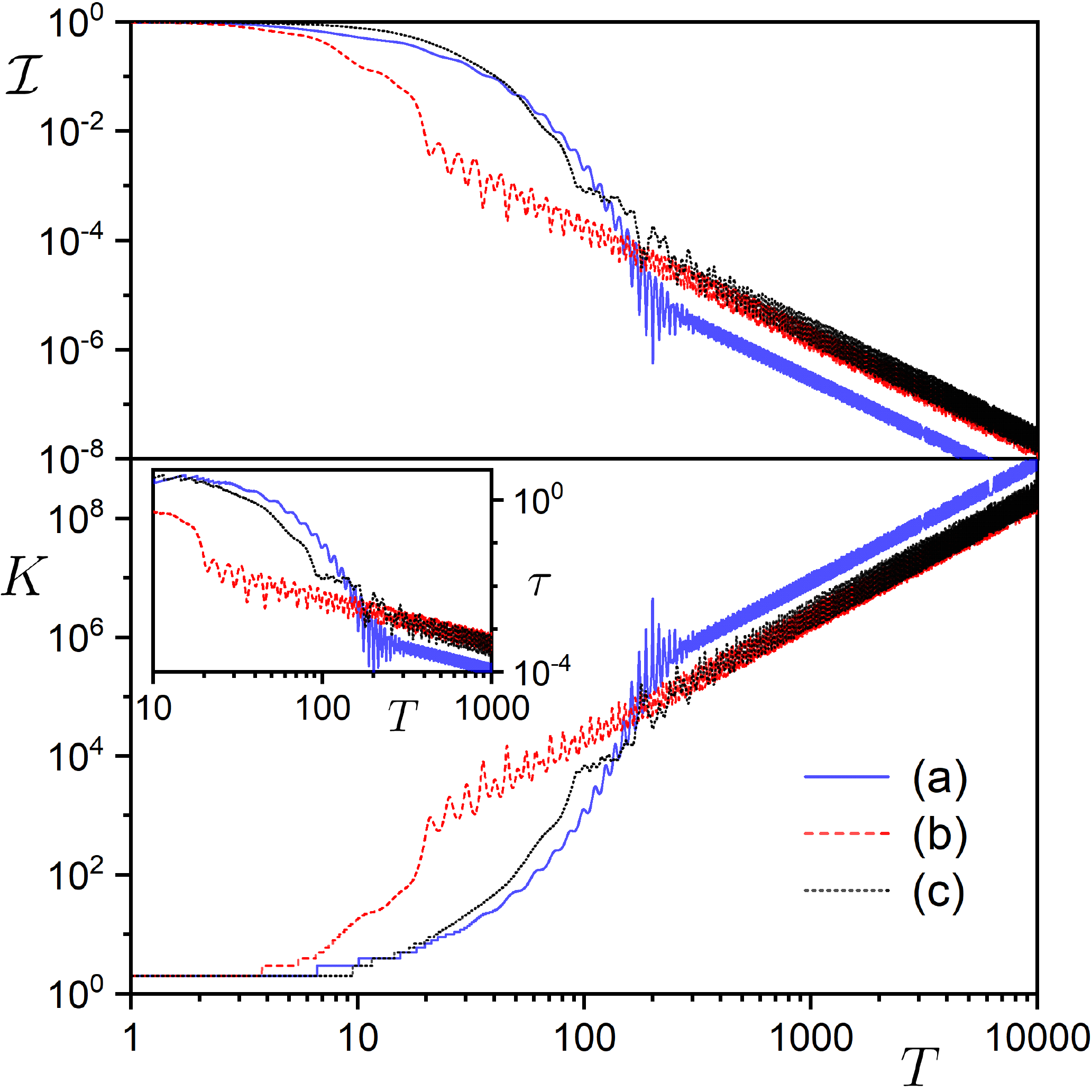}
	\caption{
Comparison of coherent and decoherence-assisted drivings along the above-described paths (a),\,(b) and (c) in ${N=10}$ Lipkin model.
Upper panel: The final infidelity~${\cal I}(T)$ as a function of time $T$ for coherent driving.
Lower panel: The minimal number of steps $K$ for decoherence-assisted protocols to get a~lower final infidelity than in the coherent driving for the given path and value of $T$.
The inset shows the corresponding value of the decoherence time ${\tau=T/K}$.  
	}
	\label{comp}
\end{figure}

How to compare these results to the performance of a fully coherent driving of an isolated system?
The upper panel of Fig.\,\ref{comp} shows the infidelity obtained in the coherent driving along the above three paths (a),\,(b) and (c) as a function of the total driving time $T$.
All curves exhibit a crossover from the exponential Landau-Zener regime \cite{Land32}, valid at smaller values of~$T$, to the asymptotic-$T$ regime characterized by the algebraic dependence ${{\cal I}(T)\propto T^{-2}}$~\cite{Orti08}.
Note that the observed oscillations of ${\cal I}(T)$ are physical.
Features of these dependencies are discussed in Ref.\,\cite{Matu22} along with the fact that the geodesic driving (a) is not generally optimal (here we observe its preference only in the asymptotic-$T$ regime).

We know that in all decoherence-assisted driving protocols, the final infidelity for any total time $T$ converges to unity if the decoherence time ${\tau=T/K}$ decreases to zero. 
The lower panel of Fig.\,\ref{comp} depicts the minimal number of steps $K$ for which the decoherence-assisted protocol along the respective path (a),\,(b) or (c) yields a better result (a lower final infidelity) than the coherent driving (the corresponding values of $\tau$ are shown in the inset).
From a practical viewpoint, the most important time domain is the one in which the coherent driving is still in the Landau-Zener regime.
In these cases the performance of the state-preparation procedure can be considerably improved by using decoherence-assisted protocols with relatively small values of $K$ (roughly from 10 to 1000).
 
{\it Conclusions.---}
We propose a new type of state preparation driving protocol based on repeated measurement-like interactions with an external system that steadily destroy coherence of the driven system in the running Hamiltonian eigenbasis.
In analogy to the quantum Zeno effect, the final fidelity converges to unity in the limit of infinite-rate interactions. 
However, even with finite-rate interactions, the new protocols can provide much better results than the corresponding coherent protocols.
Our analysis shows that the design of an optimal decoherence-assisted protocol relies on the geometric approach to quantum systems in terms of the Provost-Vallee metric.
In particular, the highest fidelity is obtained if the discretized driving procedure in the parameter space follows the geodesic path with a constant speed on the ground-state manifold.
This is in contrast to the coherent case for which geodesic driving does not play any exceptional role.
We believe that these findings will have practical applications in quantum information technologies.

{\it Acknowledgment.---}
We acknowledge financial support of the Czech Science Foundation (Grant No. 20-09998S) and the Charles
University in Prague (UNCE/SCI/013).


\end{document}